\documentstyle[preprint,pre,aps,eqsecnum]{revtex}
\begin{document}
\preprint{PISA IFUP-TH XX, June 1993}
\draft
\title{
The Kramers equation simulation algorithm:\\
I. Operator analysis
}
\author{Matteo Beccaria$^{(1,2)}$, Giuseppe Curci$^{(2,1)}$}
\address{
(1) Dipartimento di Fisica, Universit\'a di Pisa\\
Piazza Torricelli 2, I-56100 Pisa, Italy\\
(2) I.N.F.N., sez. di Pisa\\
Via Livornese 582/a, I-56010 S. Piero a Grado (Pisa) Italy
}
\date{\today}
\maketitle
\begin{abstract}
Using an operatorial formalism, we study the Kramers equation
and its applications to numerical simulations.
We obtain classes of algorithms which
may be made precise at every desired order in the time step $\epsilon$
and with a set of free parameters which can be used to reduce
autocorrelations.
We show that it is possible to use a global Metropolis test to restore
Detailed Balance.
\end{abstract}
\vskip 1truecm
\pacs{PACS numbers: 11.15.Ha, 05.50.+q, 02.50.+s}
\section{Introduction}
Quantum field theories can be numerically simulated on a lattice to obtain non
perturbative informations. A definite continuum limit
not always exists, but for many interesting theories asymptotic
freedom allows for the extraction of continuum physics from the lattice.
The algorithmic problem of the Monte Carlo approach amounts to generate
many statistically
independent configurations of the fields. These configurations must be
distributed according to a definite weight which depends on the specific
action of the model.
We have a great freedom in building such a procedure. A possible strategy
is based on the idea
of obtaining the final target distribution via a diffusion process in the
configuration space which  gives asymptotically infinite configurations
properly
distributed.
A numerical simulation of the above continuous diffusion process leads
naturally to
approximated schemes for the generation of the equilibrium ensemble.
Indeed, solving the diffusion process is
like integrating a differential equation and some kind of extrapolation
in the integration step
must be done at the end to get the
exact results \footnote{
In~\cite{beccaria} we have studied a different implementation of the
diffusion process. A diffusive
process can be realized on a discrete space-time. The correct diffusion
equation in the continuum
is obtained only at the end with a careful limit in both the spatial and
temporal steps.
Ref.~\cite{beccaria} showed the applicability and power of this method in
the framework of the
First-Exit-Mean-Time problems. Space becomes discrete and at every sweep
the degrees of freedom $q_n$
are modified to $q_{n+1} = q_n+\Delta_{\alpha_n} e_{\alpha_n}$
where $\{e_\alpha\}$ identifies all the possible
directions on the lattice and $\Delta_\alpha$ is the space step in the
$\alpha$-th direction.
The relevant direction $\alpha$ is chosen with a definite space-dependent
probability $P_\alpha$
such that as $t\to+\infty$ with $\Delta_\alpha\to 0$ a definite distribution
for the
configurations in $\{q_n\}$ is obtained.}.
On the other hand, exact algorithms may be obtained with the introduction
of a Metropolis test, namely a global recall of the action that we are
simulating.
Depending on the problem under
study one of the two choices may be preferable.
In this work we show how one can generalize the simplest Langevin process
to more complicated and
efficient schemes.
We also show which global correction must be performed in order to restore
the Detailed Balance
condition.

In Sec.~\ref{simulations} we recall the basic ideas of the Monte Carlo
approach to
numerical simulations.
In Sec.~\ref{fp} we introduce formalism and notation discussing
the Langevin and the Horowitz algorithms.
In Sec.~\ref{general} we show their natural
extensions and discuss in Sec.~\ref{correlations} their autocorrelations
in the continuum.
In Sec.~\ref{exact} we show exact versions of these schemes.
Some comment are collected in Sec.~\ref{discussion} on the relations
with existing algorithms.
Finally, Sec.~\ref{summary} is devoted to the conclusions. In
Appendix~\ref{app:symplectic}
we illustrate the symplectic methods for the integration of the equations
of motion.
Finally, in Appendix~\ref{app1} and~\ref{app2} some technical notes are
collected.
\section{Monte Carlo simulations}
\label{simulations}
Let us consider $R^n$ flat space with Euclidean measure. Given the
degrees of freedom
$\{q\}$ and the action $S(q)$ (a scalar function of the $q$) we want to
generate efficiently samples of statistically independent configurations
distributed according to the weight
\begin{equation}
P(q) = e^{-\beta S(q)},
\end{equation}
where $\beta$ is the inverse temperature of the system under study.
The strategy of Monte Carlo simulations is to start from a random initial
configuration and to evolve it along a discrete Markov chain determined by
a given transition probability
\begin{equation}
P(q^{'} \to q^{''}) \qquad\qquad \int P(q^{'}\to q^{''})\ dq^{''} = 1
\end{equation}
which is the probability of making a transition from $q^{'}$ to $q^{''}$.
We know~\cite{sokal}
that if the above Markov chain is irreducible and a stationary state $W(q)$
exists satisfying the {\em stationarity condition}
\begin{equation}
\label{stationarity}
\int W(q^{'}) P(q^{'}\to q^{''})\ dq^{'} = W(q^{''}),
\end{equation}
then the process will converge to $W(q)$. A sufficient but not necessary
condition to
satisfy Eq.~(\ref{stationarity}) is the stronger {\em Detailed Balance}
condition
\begin{equation}
W(q^{'}) P(q^{'}\to q^{''}) = W(q^{''}) P(q^{''}\to q^{'}).
\end{equation}
More generally, a powerful trick is to make the action depend on some
auxiliary
variables $\xi$ which diffuse together with the $q$ and satisfy
\begin{equation}
S(q) \to \tilde{S}(q,\xi)\quad\mbox{such that}\quad
\int d\xi e^{\tilde{S}(q,\xi)}
\sim e^{S(q)}.
\end{equation}
Averages of functionals independent of $\xi$ are equal to
those obtained with $S(q)$.
Detailed Balance will be encoded by a generalized
Eq.~(\ref{stationarity}) of the form
\begin{equation}
W(q^{'},\xi) P(q^{'},\xi^{'}\to q^{''},\xi^{''}) =
W(q^{''},T\xi^{''}) P(q^{''},T\xi^{''}\to q^{'},T\xi^{'}),
\end{equation}
where $T\xi$ is the transformed of $\xi$ under time reversal.

Choosing the best transition probability depends on
the nature of the problem under study and on which kind of measurements
one is interested in.
Moreover, for practical implementations, it is relevant to know the
specific architecture of the
machine which one will use.
As we have explained in the introduction, a
natural idea is to take advantage of a continuum dynamics towards
equilibrium.
This leads us to consider the Fokker-Planck diffusion equation
\cite{risken}, the Langevin algorithm and its generalizations.
\section{Fokker-Planck evolution}
\label{fp}
The standard Fokker-Planck equation for the diffusion in $R^n$ may be written
\begin{equation}
\label{fokkerplanck}
-\frac{\partial P}{\partial t} = H P,
\end{equation}
where the evolution kernel for the time dependent probability distribution
$P(q,t)$ is
\begin{equation}
H = \frac{1}{\beta}\ \Pi^2 + i\Pi F(q)\qquad\qquad \Pi =
-i\frac{\partial}{\partial q}
\end{equation}
and $F(q)$ is the driving force which we assume derived from a potential
\begin{equation}
F(q) = -i\Pi S(q).
\end{equation}
Eq.~(\ref{fokkerplanck}) describes relaxation towards the static distribution
\begin{equation}
W(q) = e^{-\beta S(q)}.
\end{equation}
This can be shown in the continuum evolution as follows~\cite{drummond}.
Consider the scalar
product (we restrict ourselves to real functions)
\begin{equation}
(f, h) = \int (dq)_W\ f(q) h(q) \qquad (dq)_W = dq\ \left[W(q)\right]^{-1} ,
\end{equation}
then we have
\begin{equation}
\label{eigenvalues}
(f, Hf) = \int (dq)_W\ f(q) \Pi\left\{\frac{1}{\beta}\Pi + i
F(q)\right\} f(q) =
-\frac{1}{\beta}\int  dq W(q) \left\{\Pi f(q) \left[W(q)\right]^{-1}
\right\}^2.
\end{equation}
$W(q)$ is an eigenstate of $H$ with zero eigenvalue. From the above
equation it follows that
if $Hf=0$ then $f\sim W(q)$ whence $W(q)$ is the unique stable mode.
Moreover from
Eq.~(\ref{eigenvalues}) we deduce that all the other eigenvalues of
$H$ are negative. Therefore,
by expanding a generic solution $u(q,t)$ of Eq.~(\ref{fokkerplanck})
in a basis of $H$ eigenvectors we show that $u\to W$ as $t\to +\infty$.

Consider now the Markov chain defined by the following real
transition probability
\begin{equation}
P(q^{'}\to q^{''}) = \langle q^{''} |
\exp\left(-\epsilon H\right) | q^{'} \rangle
\stackrel{def}{=} \langle q^{''} | K(q) | q^{'} \rangle,
\end{equation}
where $\epsilon$ is an auxiliary simulation time step.

A necessary condition for $W(q)$ to be a stationary distribution of
this process is
the {\em differential} equation
\begin{equation}
K(q) W(q) = 0.
\end{equation}
On the other hand, Detailed Balance may be written as the {\em operatorial}
equation
\begin{equation}
K(q) W(q) = W(q) K^\dagger(q).
\end{equation}
Of course we cannot evaluate exactly the matrix elements of $K$. This would be
equivalent to solving exactly the quantum mechanical problem
associated to the path-integral quantum extension of $S(q)$.

Many algorithms arise in trying to approximate the kernel $K$.
We consider now some well known proposals. The basic idea is to factorize
approximately the kernel
$K$ and to read the factorization as a succession of elementary steps in
the update of the old
Markov state.
\subsection{Langevin algorithm}
The standard Langevin algorithm splits $H = H_1 + H_2$
with
\begin{eqnarray}
H_1 &=& \frac{1}{\beta}\ \Pi^2, \\
H_2 &=& i\Pi F(q). \nonumber
\end{eqnarray}
Then
\begin{equation}
P(q^{'}\to q^{''}) = \int d\xi\ \langle q^{''} | e^{-\epsilon H_1} |
\xi \rangle
\langle \xi | e^{-\epsilon H_2} | q^{'} \rangle + O(\epsilon^2)
\end{equation}
The sum over $\xi$ can be read as the composition of two successive updates
realized with each of the two operators $H_1$ and $H_2$. They have trivial
matrix
elements
in the coordinate representation~\cite{chin}.
\begin{eqnarray}
\label{transition}
\langle q^{''} | e^{-\epsilon H_1} | q^{'}\rangle &=&
\left(\frac{\beta}{2\pi\epsilon}\right)^{n/2}
\exp\left[
-\frac{\beta}{2\epsilon} (q^{''}-q^{'})^2
\right],\\
\langle q^{''} | e^{-\epsilon H_2} | q^{'} \rangle &=&
\delta(q^{''}-q(\epsilon))
\end{eqnarray}
where $q(t)$ is defined by
\begin{equation}
q(0) = q^{'} \qquad \dot{q}(t) = F(q).
\end{equation}
The resulting explicit update is
\begin{equation}
q^{''} = q(\epsilon) + \sqrt{\frac{2\epsilon}{\beta}}\  \xi,
\end{equation}
where $\xi$ is a Gaussian random number with zero mean and unit variance.

This procedure introduces a systematic error in the construction of the
statistical sample since the static distribution is
now the zero mode of the operator $\bar{H}$ defined by
\begin{equation}
\exp\left(-\epsilon H_1\right) \exp\left(-\epsilon H_2\right) =
\exp\left(-\epsilon \bar{H}\right)\qquad\qquad \bar{H} = H +
\epsilon\delta_1 H + \dots .
\end{equation}
Solving perturbatively in $\epsilon$ the equation
\begin{equation}
(H+\epsilon\delta_1 H + \dots) (W + \epsilon\delta_1 W + \dots ) = 0
\end{equation}
we find a non trivial $\delta_1 W$ which affects
to order $O(\epsilon)$
all the averages on the sample
\begin{equation}
W + \epsilon \delta_1 W = \exp\left\{-\beta\left[S + \frac{\epsilon}{4}
\left(\nabla^2 S - \beta
\left(\nabla S\right)^2\right)\right]\right\}.
\end{equation}
Therefore,
Detailed Balance holds exactly only with respect to $\bar{H}$.
Of course, we are not able to build explicitly the function $q(t)$. At
this order we can
use
\begin{equation}
q^{''} = q^{'} + \epsilon F(q^{'}) + \sqrt{\frac{2\epsilon}{\beta}}\  \xi
\end{equation}
and the only consequence is a different explicit form of the correction
term which becomes
now
\begin{equation}
W + \epsilon \delta_1 W^\prime = \exp\left\{-\beta\left[S +
\frac{\epsilon}{4}\left(\nabla^2 S -
\frac{\beta}{2}
\left(\nabla S\right)^2\right)\right]\right\}.
\end{equation}
\subsubsection{Smart Monte Carlo}
In this section we show how the structure of the Detailed Balance violations
in the Langevin algorithm (which the Smart Monte Carlo method~\cite{rossky}
tries to correct)
suggests the extension of the dynamics in an enlarged $(q,p)$
space.
In this wider space the diffusive process modifies some approximation of
a canonical Hamiltonian evolution.
The Langevin update is ($\epsilon\to\epsilon^2/2$)
\begin{equation}
q^{'} = q + \frac{\epsilon^2}{2} F(q) + \epsilon\ \xi
\end{equation}
corresponding to the transition probability in Eq.~(\ref{transition}).
The interesting point
is that the parameter controlling Detailed balance violations
\begin{equation}
x = \frac{W(q^{'}) \langle q^{''} | K(q) | q^{'}\rangle}{W(q^{''})
\langle q^{'} | K(q) | q^{''}\rangle}
\end{equation}
may be written
\begin{equation}
x = \exp\left(H(q^{''}, p^{''})-H(q^{'}, p^{'})\right),
\end{equation}
where $H(q,p) = S(q) + \frac{1}{2} p^2$ and
\begin{eqnarray}
p^{'} &=& \xi, \nonumber \\
q^{''} &=& q^{'} + \frac{\epsilon^2}{2} F(q^{'}) + \epsilon p^{'},\\
p^{''} &=& p^{'} + \frac{\epsilon}{2}\left(F(q^{'}) + F(q^{''})\right),
\nonumber
\end{eqnarray}
and this set of equations can be recast in the form of a standard leap-frog
symplectic
integrator (see Appendix~\ref{app:symplectic})
\begin{eqnarray}
p^{'} &=& \xi, \nonumber \\
\bar{p}\ &=& p^{'} + \frac{\epsilon}{2} F(q^{'}), \nonumber \\
q^{''} &=& q^{'} + \epsilon \bar{p}, \\
p^{''} &=& \bar{p} + \frac{\epsilon}{2} F(q^{''}). \nonumber
\end{eqnarray}
The conclusion is that the equality $x=1$ can be enforced improving the
approximation
of the evolution determined by $H$.
\subsection{Hyperbolic algorithm}
This is a clever proposal~\cite{horowitz1} which improves the naive
Langevin algorithm
using the dynamics of the Kramers equation~\cite{kramers,risken}.
We double the degrees of freedom $q\to (q,p)$ and
introduce the operator
\begin{equation}
H = \frac{\gamma}{\beta}\  \Pi_p^2 + i\Pi_p (-\gamma p) + i \Pi_p F(q) +
i \Pi_q p,
\end{equation}
where $\gamma$ is an arbitrary parameter.

The proof of convergence to equilibrium given for the Fokker-Planck
evolution cannot be applied.
However, following~\cite{horowitz1}, $H$ determines a Hamiltonian
flow in $(q,p)$ space with
a diffusion term in the $p$ variables. Hence if a solution of
$\dot{u} = Hu$ is positive at $t=0$,
it remains positive at any time later as can also be seen from the
representation of the
time evolution of $H$ in terms of a path-integral with positive measure.
On the other hand,
integrating by parts any eigenvector $u_n$ with non zero eigenvalue we get
\begin{equation}
\int dq\ dp\ u_n = 0
\end{equation}
and assuming some eigenvalue with positive real part we contradict
positivity of the evolution
determined by $H$.

We introduce an $O(\epsilon)$ error by splitting $H = H_1+H_2$ with
\begin{eqnarray}
H_1 &=& \frac{\gamma}{\beta}\  \Pi_p^2 + i\Pi_p (-\gamma p) +
i \Pi_p F(q), \\
H_2 &=& i \Pi_q p \nonumber
\end{eqnarray}
we can write down an approximated update for $H$ obtained from $H_1$
and $H_2$
\begin{eqnarray}
p^{''} &=& (1-\epsilon\gamma) p^{'} +\epsilon F(q^{'}) +
\sqrt{\frac{2\epsilon\gamma}{\beta}}\ \xi^\prime, \\
q^{''} &=& q^{'} + \epsilon p^{''}.
\end{eqnarray}
The equilibrium distribution is
\begin{equation}
W(q,p) = \exp\left\{-\beta\left(\frac{1}{2} p^2 + S(q) +
\epsilon S_1(q,p,\gamma)+\cdots\right)\right\}.
\end{equation}
The parameter $\gamma$ can be tuned to minimize the
autocorrelation time of the observable which we are going to measure.
Unlike the Langevin case, it is impossible to
give an expression of $S_1(q,p,\gamma)$ in the form of a polynomial
in $S(q)$ and its derivatives.

On the basis of the above general arguments we should have introduced a
systematic error
of order $O(\epsilon)$ due to $S_1(q,p,\gamma)$. This holds true for
averages of
generic functions $\Omega(q,p)$ which depend on the auxiliary variables.
However, if we are interested in functions $\Omega(q)$ of $q$ only,
it is shown
in~\cite{horowitz1}
that the particular form of $S_1(q,p,\gamma)$ reduces the error to
$O(\epsilon^2)$.

Assuming the $q$ even under time reversal, the $p$ variables must be are odd
under $T$. This implies that
the Detailed Balance condition must be written~\cite{risken}
\begin{equation}
K(q,p) W(q,p) = W(q,p) K^\dagger(q,-p)
\end{equation}
and in our case it is violated by $O(\epsilon)$ terms because of
$S_1(q,p,\gamma)$.

An interesting variant of this procedure is called Guided Hybrid
Monte Carlo~\cite{horowitz2},
this is an algorithm which admits an exact extension and we shall
discuss it later as a
particular case of our methods.
\section{Improved algorithm}
\label{general}
In this section we shall describe our approach which allows us for an
improvement of the
precision up to any desired order and which is a natural generalization of the
previous schemes.

The key point of our method is a different operator splitting of $H$.
Indeed, we can separate $H = H_{irr} + H_{rev}$ where
\begin{eqnarray}
H_{irr} &=& \frac{\gamma}{\beta}\ \Pi_p^2 + i\Pi_p (-\gamma p), \\
H_{rev} &=& i \Pi_p F(q) + i \Pi_q p.
\end{eqnarray}
The important point is that both $H_{irr}$ and $H_{rev}$
annihilate the static distribution
\begin{equation}
W(q,p) = \exp\left\{-\beta\left(\frac{1}{2} p^2 + S(q) \right)\right\}.
\end{equation}
After all, this is why the Kramers equation works in the continuum limit.
Writing
\begin{equation}
\label{unsymmetric}
\exp\left(-\epsilon H_{irr}\right) \exp\left(-\epsilon H_{rev}\right) =
\exp\left(-\epsilon\left(H_{irr} +
H_{rev} + \epsilon X\right)\right),
\end{equation}
the operator $X$ is non trivial but has the useful property $X W(q,p) = 0$.
This observation has the important consequence that
we can consistently set to zero all the corrections in powers of $\epsilon$
to $W(q,p)$ and treat independently the irreversible and the reversible
evolutions.

We expect Detailed Balance to hold exactly: indeed, as we have
remarked above, the
two kernels
\begin{equation}
\exp\left(-\epsilon H_{irr}\right)\qquad \exp\left(-\epsilon H_{rev}\right)
\end{equation}
leave invariant $W$ which is therefore a steady state for the evolution
determined by any string
$K$ of the form
\begin{equation}
K = \prod_i \exp\left(-a_i\epsilon H_{irr}\right)\
\exp\left(-b_i\epsilon H_{rev}\right),
\end{equation}
where $a_i$ and $b_i$ are constants.

Indeed, we can check that separately
\begin{eqnarray}
H_{irr}(q,p) W(q,p) = W(q,-p) H_{irr}^\dagger(q,-p), \\
H_{rev}(q,p) W(q,p) = W(q,-p) H_{rev}^\dagger(q,-p), \\
\end{eqnarray}
we conclude that the Detailed Balance condition is exactly
satisfied by a symmetric splitting like for instance
\begin{equation}
K = \exp\left(-\frac{1}{2}\epsilon H_{irr}\right)
\exp\left(-\epsilon H_{rev}\right)
\exp\left(-\frac{1}{2} \epsilon H_{irr}\right).
\end{equation}
Of course, we must specify how to evaluate the matrix elements
of the two kernels, but this is not a problem.
The irreversible part can be solved exactly.
Writing
\begin{eqnarray}
\Omega_1 &=& \frac{\gamma}{\beta}\ \Pi_p^2,\\
\Omega_2 &=& i\Pi_p (-\gamma p),
\end{eqnarray}
then we obtain
\begin{equation}
[\Omega_1 , \Omega_2] = -2\gamma \Omega_1
\end{equation}
and by the Campbell-Baker-Hausdoff formula
\begin{equation}
\exp\left(-\epsilon (\Omega_1 + \Omega_2)\right) =
\exp\left(-h(\epsilon) \Omega_1\right) \exp\left(-\epsilon \Omega_2\right).
\end{equation}
Differentiating with respect to $\epsilon$ with the condition $h(0) = 0$
we get
\begin{equation}
h(\epsilon) = \frac{1}{2\gamma} \left(1-e^{-2\gamma\epsilon}\right).
\end{equation}
The matrix element $\langle p^{''},q^{'} | p^{'},q^{'} \rangle_{irr}$
corresponds exactly to the discrete
update
\begin{equation}
p^{''} = e^{-\gamma \epsilon} p^{'} +
\sqrt{\frac{1-e^{-2\gamma\epsilon}}{\beta}}\
\xi^\prime .
\end{equation}
The reversible matrix element $\langle p^{''}, q^{''} | p^{'}, q^{'}
\rangle_{rev}$
must be some approximate integration of the deterministic Cauchy problem
\begin{equation}
\dot{q} = p\qquad \dot{p} = F(q)
\end{equation}
corresponding to the update
\begin{equation}
q^{''} = Q(q^{'},p^{'},\epsilon) \qquad p^{''} = P(q^{'},p^{'},\epsilon),
\end{equation}
where the maps $Q$ and $P$ can be, for instance, an higher order symplectic
integrator.
We stress that at this level of the discussion we do not need using a
canonical integrator
conserving the Poincar\'e invariants, in particular the phase-space measure.
The importance of symplectic integrators will be clear
in the discussion on the exact extensions of these algorithms.

We remark that the systematic error we are introducing is just the
error made in
approximating this reversible step. Its improvement is a well known problem
which can be solved up to an arbitrary precision with increasing
computational cost.

As in the hyperbolic algorithm, the free parameter $\gamma$ can be
tuned minimizing
autocorrelations times, whle the integration step $\epsilon$ must be
kept small enough
to control systematic errors.
\subsection{Extended schemes}
Following the above ideas, we can introduce schemes in which other
free parameters are present. They can be tuned to obtain even smaller
autocorrelations times.
The simplest example is
\begin{equation}
H = H_{irr} + H_{rev} + H_{rot},
\end{equation}
where
\begin{eqnarray}
H_{irr} &=& \frac{\gamma}{\beta}\ \Pi_y^2 + i\Pi_y (-\gamma y), \\
H_{rev} &=& i \Pi_p F(q) + i \Pi_q p, \\
H_{rot} &=& i\alpha \left( y\Pi_p - p\Pi_y\right).
\end{eqnarray}
The operator $H_{rot}$ is just a rotation in the $(y,p)$ plane and its
exact matrix
elements
\begin{equation}
\langle p^{''} y^{''} | \exp\left(-\epsilon H_{rot}\right) | p^{'}y^{'}
\rangle
\end{equation}
corresponds therefore to the update
\begin{eqnarray}
p^{''} &=& p^{'} \cos(\alpha\epsilon) + y^{'}\sin(\alpha\epsilon), \\
y^{''} &=& y^{'} \cos(\alpha\epsilon) - p^{'}\sin(\alpha\epsilon).\nonumber
\end{eqnarray}
A symmetric kernel like
\begin{equation}
\exp\left(-\epsilon H_{irr}\right)\exp\left(-\epsilon H_{rot}\right)
\exp\left(-\epsilon H_{rev}\right)
\exp\left(-\epsilon H_{rot}\right)\exp\left(-\epsilon H_{irr}\right)
\end{equation}
satisfies exactly the Detailed Balance condition with respect to the
equilibrium
distribution
\begin{equation}
W(q,p,y) = e^{-\beta (\frac{1}{2}(p^2+y^2)+S(q))}.
\end{equation}
The parameters $\gamma$ and $\alpha$ must be tuned, but we are unable to
provide a general rule for their choice. However we shall show that the
additional
$y$ variables improve the optimal autocorrelation of the algorithm.

An interesting feature of this scheme is that if we let $y$ reach
equilibrium, inserting
one $\exp\left(-\epsilon H_{irr}\right)$ factor near
$\exp\left(-\epsilon H_{rev}\right)$
and setting $\gamma\to\infty$, then we obtain
\begin{eqnarray}
y^\prime &=& \frac{1}{\sqrt{\beta}} \xi, \\
p^\prime &=& p\cos(\alpha\epsilon) + \frac{1}{\sqrt{\beta}}
\xi \sin(\alpha\epsilon) \nonumber
\end{eqnarray}
which is just the mixing which we find in the stochastic step of the
Guided Hybrid
algorithm. It is clear that, at least formally, we reduce to this
update starting from
our basic $(q,p)$ scheme and relating $\gamma$ to $\alpha$ writing
\begin{equation}
\exp(-\gamma\epsilon) = \cos(\alpha\epsilon) \qquad\qquad
0 < \alpha\epsilon < \frac{\pi}{2}
\end{equation}
(the case in which the cosine is negative is handled easily by a
slight generalization introduced in
the following section).

A very natural extension is obtained building in the space
$(q, p, y_1, \cdots, y_N)$ a chain of rotation operators
acting in the space of the auxiliary variables
\begin{eqnarray}
H_{rev} &=& i \Pi_p F(q) + i \Pi_q p, \\
H_{rot, 1} &=& i\theta_1 \left( y_1\Pi_p - p\Pi_{y_1}\right), \\
&& \cdots \\
H_{rot, N} &=& i\theta_N \left( y_N\Pi_{y_{N-1}} -
y_{N-1}\Pi_{y_N}\right), \\
H_{irr} &=& \frac{\gamma}{\beta}\ \Pi_{y_N}^2 + i\Pi_{y_N} (-\gamma y_N),
\end{eqnarray}
where the angles $\theta_i$ are free.
\subsection{Some remarks}
\subsubsection{Arbitrary rotations}
\label{stress}
The reason why it is so natural to introduce the rotation
operators $H_{rot}$ is the following:
if $p$ be a $N$-component vector, then the discrete update
\begin{equation}
p^{''} = x C p^{'} + \sqrt{1-x^2} \xi
\qquad C\in{\rm O(N, {\rm I\hskip -2pt{\sf R}})}
\end{equation}
generates a normally distributed succession
$\{p\}$: $P(p) = \exp\frac{1}{2}p^2$.

The connected component of this update may be written in terms of the
exponential of $\epsilon$~times a differential operator by introducing
all the generators of
${\rm so(N, {\rm I\hskip -2pt{\sf R}})}$
\begin{equation}
H_{rot} = i\ \Theta_{ij}\ p_i\ \Pi_{p_j}\qquad \Theta_{ij} = -\Theta_{ji}.
\end{equation}
The extended scheme of the previous section is just a particular
choice of $C$ realized as a
chain of elementary rotations.

In the steps where the matrix element of the transition kernel
are computed exactly,
the parameter $\epsilon$ does not control an approximation any more.
Indeed, our convergence analysis deals with the limit $\epsilon\to 0$
irrespectively of $\gamma\epsilon$, $\Theta_{ij}\epsilon$.
On a formal level,
one can take parameters $\gamma$ or $\Theta_{ij}$ not $O(1)$
with respect to $\epsilon$. From
a practical point of view this means that we can choose $C$
independent of $\epsilon$;
intuitively when the dynamics is switched off one is left with a
free rotation in
the $p$ space\footnote{
An interesting example of this mechanism if the following. We
could try looking for an extension
of the basic update
\begin{equation}
p^{''} = \alpha p^{'} + \beta \xi
\end{equation}
to the case $\alpha\in \sf C$ and $\beta$ to be determined. However,
by complexifying
$p$ and $\xi$ representing them as 2-vectors
\begin{equation}
p = u+iv\qquad \xi = \eta + i p^{''}i
\end{equation}
we have
\begin{equation}
p^{''} = |\alpha| C p^{'} + \beta \xi \qquad C =
\frac{1}{|\alpha|}\left({\rm Re}\alpha - i\sigma_2
{\rm Im}\alpha\right)\in{\rm O(2, {\rm I\hskip -2pt{\sf R}})}
\qquad\mbox{$\sigma_2$ is a Pauli matrix}
\end{equation}
and choosing $\beta = \sqrt{1-|\alpha^2|}$ we fall in the above case,
since $C$ may have a limit
different from 1 as $|\alpha| \to 1$.
}.

The vector $p$ may be the full set of generalized momenta in a
single site. However, it can also mix momenta from different sites.
This introduces long distance
reshufflings and allows for pair-exchanges which are represented by
orthogonal matrices.
\subsubsection{Random parameters}
In the Hybrid Monte Carlo~\cite{hmc1,hmc2}, better autocorrelations
may be obtained by randomizing the
trajectory length. Here, a probability distribution for parameters
like $\gamma$ may be introduced.
However, from preliminar numerical investigations we have indications
that this device
should not be a major improvement.
\subsubsection{Dynamical Fermions}
Consider a model with dynamical fermions. As is well known, in such a
situation, the most
expensive part of a lattice simulation is the inversion of the fermionic
propagator.
The time spent during this step becomes the natural unit of time.
A fundamental parameter is then the number of inversions per sweep which
we shall call $\cal R$.
A common implementation of dynamical fermions~\cite{hmc1} replaces the
fermionic degrees of freedom
with bosonic fields $\chi$ which are updated by a heat-bath step.
The refreshment frequency is relevant here. As an example, let us
integrate the equations of
motion by the simplest leap-frog scheme of Eq.~(\ref{simplest}). If
we refresh the $\chi$ fields
every $k$ sweeps and if $N_{md}$ is the number of molecular dynamics
steps in each sweep, then
\begin{equation}
{\cal R } = N_{md} + \frac{1}{k}.
\end{equation}
In the Hybrid Monte Carlo one usually takes $N_{md}\gg 1$ and $k=1$. In
the schemes we have discussed
one has always $N_{md} = 1$ so that taking $k = 1$ is twice slower than a
large $k$.
\section{Discussion of correlations}
\label{correlations}
Consider a function $\Omega(q)$ of the dynamical variables.
The evolution in time of the sample average of $\Omega$ is given by
\begin{equation}
\frac{\partial}{\partial t} \Omega = -H^\dagger \Omega .
\end{equation}
Consider the continuum limit in which $\epsilon\to 0$ and let us compare
the evolution
of $\langle q\rangle $ in the free theory where
\begin{equation}
S(q) = \frac{1}{2} \omega^2 q^2
\end{equation}
for the three cases discussed above.
We restrict the discussion to the case in which all the tunable parameters
are independent
of $\epsilon$ (see the discussion in paragraph~(\ref{stress}).
We have explicitly
\begin{eqnarray}
-H_{lang}^\dagger &=& \frac{1}{\beta}\ \partial_q^2 - \omega^2 q
\partial_q, \nonumber\\
-H_{hyp}^\dagger &=& \frac{1}{\beta}\ \partial_p^2 - \gamma p \partial_p
-\omega^2 q \partial_p + p\partial_q,\\
-H_{(q,p,y)}^\dagger &=& \frac{1}{\beta}\ \partial_{y_N}^2 - \gamma y_N
\partial_{y_N}
+\sum_{i=1}^N \theta_i\left(y_i\partial_{y_{i-1}} - y_{i-1}\partial_{y_i}
\right)
-\omega^2 q \partial_p + p\partial_q, \nonumber \\
&& (y_0 = p). \nonumber
\end{eqnarray}
The eigenvalues of the time evolution of $\langle q\rangle $ give directly
the
intrinsic autocorrelation of the Markov chain in the $\epsilon\to 0$ limit.

For the Langevin algorithm, we have a decay with eigenvalue
$\lambda = -\omega^2$
and no possibility of tuning any parameter.

In the hyperbolic case, we have two eigenvalues $\lambda_\pm$ and
\begin{equation}
\min_\gamma\max ( \lambda_+, \lambda_- ) = -\omega
\end{equation}
obtained at $\gamma=2\omega$. Dimensions are different with respect to
Langevin,
because the Kramers equation is second order; afterall,
this is why the hyperbolic algorithm is better than
the Langevin one.

In the $(q,p,y)$ extended scheme the roots of the secular equation
$\lambda_i$ are tedious functions of $\theta_i$ and $\gamma$. We
remark that a possible choice is to choose $\gamma$ and $\theta_i$ in
such a way
to have the maximum possible degeneracy in $\lambda$.
Some explicit results for low values of $N$ are shown in
Tab.~\ref{solutions} where all the variables has been made adimensional
dividing out by $\omega$.
One can improve the autocorrelation of $\langle q\rangle$ of at least
a factor $\sqrt{11}$ in the $N=5$ case.
It seems reasonable that this mechanism can be improved with no limit
introducing more and more auxiliary variables
(up to $N<6$ Tab.~\ref{solutions} satisfies
$\lambda_N = -\sqrt{2N+1}$)
and we must balance the increasing computational work needed to
tune all the associated new parameters and the gain in autocorrelation
of the sample.

If we take into account the details of the discrete dynamics at
$\epsilon > 0$ we obtain linear
recursive equations giving the evolution of the momenta
instead of differential equations. Their characteristic polynomials
determine the $\epsilon$ dependent relaxation eigenvalues. For
$\epsilon$ small enough the
qualitative picture which one finds in the continuum does not change.

Of course, we cannot be too optimistic. In a realistic model there
will be many relevant
frequencies. An optimal tuning for all of them will average on the their
separate behaviours. Since we cannot give analytical estimates for
autocorrelations in
interacting models, explicit numerical tests are needed.
\subsection{Curved Space}
We are mainly interested in the generalization from flat space to
the case of the curved manifold
of a Lie group $G$. We must endow the configuration space with a
definite metric and
write down the corresponding covariant Fokker-Planck or Kramers equation.
The most convenient and natural choice is the left and right invariant metric
$ds^2 = \mbox{Tr}(dU\ dU^\dagger)$. Its Laplace-Beltrami inavriant
operator is built by substituting
derivatives $\partial_\alpha$ with left-derivatives on the group
$\nabla_\alpha$ defined by
\begin{equation}
f(\exp(i\epsilon^\alpha T^\alpha) g) = f(g) +
\epsilon^\alpha\nabla_\alpha f(g) +
O(\epsilon^2).
\end{equation}
This choice is convenient because $\partial_\alpha$ was the generator
of translations in flat
space whereas on the group $\nabla_\alpha$ is simply the generator
of translations by
left multiplication
\footnote{We recall that dealing with $\nabla_\alpha$
we can integrate by parts with respect to the Haar measure $d\mu(g)$.
}.
In the Langevin case, the diffusive term is the heat kernel
\begin{equation}
\langle g^\prime | e^{-\epsilon \nabla^2} | g^{\prime\prime}\rangle =
\langle 1 | e^{-\epsilon \nabla^2} | \left(g^\prime\right)^{-1}
g^{\prime\prime}\rangle ,
\end{equation}
namely the matrix elements of the Laplace-Beltrami operator on $G$.
For $SU(N)$ a compact expression
can be given~\cite{menotti} which however is gaussian in the Lie
parameters only if $g^\prime$
is near to $g^{''}$ (the weak coupling limit of $QCD$). The advantage
of the Kramers equation is
that it is associated to a diffusion on $R^n\times G$ where $n = \dim G$,
the $p$ variables
being flat. The molecular dynamics on the group is more
complicated than in the flat case, but no curved heat kernel is needed
here because the Hamiltonian
evolution on $G$ is modified by a diffusion in its flat Lie algebra.
\section{Exact Extensions}
\label{exact}
Until now we have considered methods which are not exact and need an
extrapolation to obtain
the exact $\epsilon\to 0$ limit. Moreover, finite $\epsilon$ effects
can be studied by the analysis
of the effects of the correction terms in the equilibrium action and
it is not always clear if
critical behaviour is unchanged if this terms are present.
Nevertheless, they have some advantages over exact methods; on a formal
level, they can deal with
complex actions and they do not suffer the decrease of efficiency with
increasing volume of
the Hybrid Monte Carlo algorithm  due to the acceptance rate.

Anyhow, adding a global accept/reject test, we can write
exact extensions which
satisfy the Detailed Balance condition.
Let us consider the minimal $(q,p)$ scheme of the improved Hyperbolic
algorithm;
let us introduce an irreversible
transition probability $\Pi$, a reversible one $T$ and an accept/reject
test associated
with the acceptance probability $A$.
We want to prove Detailed Balance in full generality so we point out
which properties of
these components we need.

The irreversible transition probability must satisfy
\begin{eqnarray}
\Pi(p^{'}\to p^{''}) &=& \Pi(-p^{'}\to -p^{''}), \\
\int  \Pi(p^{'}\to p^{''})\ dp^{''} &=& 1, \\
W(q,p^{'}) \Pi(p^{'}\to p^{''}) &=& W(q,p^{''}) \Pi(p^{''}\to p^{'}) ,
\end{eqnarray}
where we have put $\beta=1$ and $W(q,p) = \exp(-S(q,p))$ is the
equilibrium state.

The properties of the reversible transition probability are
\begin{eqnarray}
\int T(q^{'},p^{'}\to q^{''},p^{''})\ dq^{''}\ dp^{''} &=& 1, \\
T(q^{'},p^{'}\to q^{''},p^{''}) &=& T(q^{''},-p^{''}\to q^{'}, -p^{'}).
\label{exchange}
\end{eqnarray}
The requirement on $A$ is the following: define
\begin{equation}
K(q^{'},p^{'}\to q^{''},p^{''}) =
A(q^{'},p^{'}\to q^{''},p^{''}) T(q^{'},p^{'}\to q^{''},p^{''}),
\end{equation}
then we want
\begin{equation}
W(q^{'},p^{'})K(q^{'},p^{'}\to q^{''},p^{''}) =
W(q^{''},p^{''})K(q^{''},-p^{''}\to q^{'},-p^{'}).
\end{equation}
An explicit choice of $\Pi$, $T$ and $A$ which satisfies all the
above conditions is
\begin{eqnarray}
\Pi(p^{'}\to p^{''}) &=& c\ \exp\left[\frac{1}{2(1-x^2)}
(p^{''}-x p^{'})^2\right], \nonumber \\
T(q^{'},p^{'}\to q^{''},p^{''}) &=& \delta(q^{''} - Q(q^{'}, p^{'}))
\delta(p^{''}-P(q^{'},p^{'})), \\
A(q^{'},p^{'}\to q^{''},p^{''}) &=&
\min\{1, W(q^{''},p^{''})/W(q^{'},p^{'})\}, \nonumber
\end{eqnarray}
where $x$ is any real number and $Q$, $P$ are the maps associated to a
symplectic integrator.
The integrator must be canonical to avoid any Jacobian in verifying
Eq.~(\ref{exchange}).

\subsection{$(q,p)$ scheme}
Our extension of the $(q,p)$ scheme is the following
\begin{eqnarray}
(q^{'},p^{'}) &\stackrel{\Pi}\rightarrow& (q^{'},p_1), \nonumber \\
(q^{'},p_1) &\stackrel{T}\rightarrow& (q_1,p_2), \nonumber \\
\mbox{accept} &:& (q^{''}, p_3) = (q_1,p_2) \ \mbox{with}
\ \ P = A(q^{'},p_1\to q_1,p_2), \\
\mbox{reject} &:& (q^{''}, p_3) = (-p_1, q^{'}) \ \mbox{with}
\ \ P = 1- A(q^{'},p_1\to q_1,p_2),\nonumber \\
(q^{''},p_3) &\stackrel{\Pi}\rightarrow& (q^{''},p^{''}). \nonumber
\end{eqnarray}
The full transition probability of the above algorithm is
\begin{eqnarray}
\lefteqn{P(q^{'},p^{'}\to q^{''},p^{''}) = } && \nonumber \\
&&
\int \Pi(p^{'}\to p_1)K(q^{'},p_1\to q^{''} , p_2)\Pi
(p_2\to p^{''}) \ dp_1\ dp_2 + \nonumber \\
&& + \int \Pi(p^{'}\to p_1)\{1-A(q^{'},p_1\to q_1 , p_2)\}
T(q^{'},p_1\to q_1 , p_2)
\times \\
&& \times \Pi(-p_1\to p^{''})\delta(q^{'}-q^{''})
\ dp_1\ dp_2\ dq_1 . \nonumber
\end{eqnarray}
Using the properties of $\Pi$, $T$ and $A$ we verify the normalization
condition
\begin{equation}
\int P(q^{'},p^{'}\to q^{''},p^{''})\ dq^{''}\ dp^{''} = 1 .
\end{equation}
In Appendix~\ref{app1} we show that
\begin{equation}
\label{db1}
W(q^{'},p^{'}) P(q^{'},p^{'}\to q^{''},p^{''}) =
W(q^{''},p^{''}) P(q^{''},-p^{''}\to q^{'},-p^{'}),
\end{equation}
hence Detailed Balance is demonstrated.
\subsection{$(q,p,y)$ scheme}
We restrict ourselves for simplicity to the $N=1$ case.
If we assume $q$ even under time reversal, it follows that also $y$ is
even whilst $p$ is odd.
We expect to find a Detailed Balance equation of the form
\begin{equation}
\label{db2}
W(q^{'},p^{'},y^{'}) P(q^{'},p^{'},y^{'}\to q^{''},p^{''},y^{''}) =
W(q^{''},p^{''},y^{''}) P(q^{''},-p^{''},y^{''}\to q^{'},-p^{'},y^{'}) .
\end{equation}
We must remember that the matrix element of $H_{rot}$ in the $(p,y)$ plane is
exact and leaves invariant the sum of squares in the action. Hence we have
\begin{equation}
W(q^{'},p^{'},y^{'}) T_{rot}(q^{'},p^{'},y^{'}\to q^{''},p^{''},y^{''}) =
W(q^{'},p^{''},y^{''}) T_{rot}(q^{'},p^{'},y^{'}\to q^{''},p^{''},y^{''}).
\end{equation}
The scheme we propose is
\begin{eqnarray}
(q^{'},p^{'},y^{'}) &\stackrel{\Pi}\rightarrow& (q^{'},p^{'},y_1),
\nonumber\\
(q^{'},p^{'},y_1) &\stackrel{T_{rot}}\rightarrow& (q^{'},p_1,y_2), \nonumber\\
(q^{'},p_1,y_2) &\stackrel{T}\rightarrow& (q_1,p_2,y_2), \nonumber\\
\mbox{accept} &:& (q^{''}, p_3) = (q_1,p_2) \ \mbox{with}
\ \ P = A(q^{'},p_1\to q_1,p_2), \\
\mbox{reject} &:& (q^{''}, p_3) = (-p_1, q^{'}) \ \mbox{with}
\ \ P = 1- A(q^{'},p_1\to q_1,p_2),\nonumber\\
(q^{''},p_3,y_2) &\stackrel{T_{rot}}\rightarrow& (q^{''},p^{''},y_3),
\nonumber\\
(q^{''},p^{''},y_3) &\stackrel{\Pi}\rightarrow& (q^{''},p^{''},y^{''}).
\nonumber
\end{eqnarray}
In Appendix~\ref{app2} we show that even for this scheme, the Detailed Balance
condition in Eq.~(\ref{db2}) holds.
\section{Comparison with Hybrid Monte Carlo}
\label{discussion}
Let us first consider the $(q,p)$ scheme. This algorithm has been
tested on compact $QED$
in~\cite{horowitz2} where its performance has been compared with
the global Hybrid Monte Carlo.

In the $\gamma\to\infty$ limit, we recover exactly the Hybrid Monte
Carlo algorithm with a single
molecular dynamics step. The quantity $(\gamma\epsilon)^{-1}$ plays
qualitatively the role of
$N_{md}$ and a finite $\gamma$ tries to reproduce a $N_{md}>1$ dynamics
by taking into account
the old momenta of the microcanonical evolution.

We remark that, after integrating out the fermions, the non local
nature of the action
prevents one from taking advantages from local Hybrid Monte Carlo
algorithms~\cite{rossi}.
The expected scaling behaviour is discussed as follows.
Assume our lattice model to satisfy $L \gg \xi$, where $L$ is the
lattice size and $\xi$
the correlation length, then the acceptance
probability of the HMC behaves as
$P_{acc} \sim \mbox{erfc}(c N_{md} \epsilon^3 \sqrt{V})$. Optimal
tuning is expected to give
$N_{md} \sim 1/\epsilon$ with a sweep-sweep correlation $\tau\sim 1/\epsilon$.
To keep the acceptance rate constant while varying the volume
needs the scaling
$\epsilon \sim V^{-1/4}$. The Kramers algorithm has $N_{md} = 1$
independently on the tuning parameters. We {\sl assume} that the
scaling relation is modified
to $\epsilon \sim V^{-1/6}$; this assumption amounts to neglect
the effects of momenta negation
on the acceptance probability. If the optimal autocorrelations
stay $\sim 1/\epsilon$
then under our assumption, the HMC is expected to behave badly on
larger volumes.
Moreover, from the point of view of numerical precision, the
opportunity of having
$N_{md} = 1$ is welcome; indeed, in $QCD$, as the volume ($L^4$) grows
or the quark mass ($m$)
is decreased we must have $N_{md} \sim L/m^{3/4}$ and some protection
seems necessary
to protect irreversibility against the accumulation of numerical errors.

The same qualitative discussion applies to the extended schemes.
In a realistic model many frequencies are present and the $\lambda$
values of Tab.~\ref{solutions}
are too optimistic. However, we expect an improvement of efficiency
with an appropriate
tuning of $\theta_i$ and the same qualitative scaling with volume.
\section{Conclusions}
\label{summary}
In summary, we have analyzed the Kramers equation approach to
lattice simulations. Our
operatorial approach is an unified framework in which the
convergence analysis is straightforward.
Generalized algorithms come out easily with an increasing
number of free parameters.
They eventually can be tuned in order to minimize autocorrelations.
Explicit numerical tests are
needed. All these algorithms are particularly safe from the point of
view of numerical
precision and they are indicated when a great deal of computational
effort is needed.
Subsequent papers will be devoted to a numerical confirmation of
the convergence and scaling
properties of these proposals. In particular, we are interested
in models with dynamical
fermions where the numerical precision problem may be important.
\acknowledgments
We thank Prof. Paolo Rossi for a careful reading of the manuscript and
many useful discussions.
\appendix
\section{Symplectic methods}
\label{app:symplectic}
In this appendix we address the problem of the numerical canonical
integration of the
Hamilton equations.
Many studies of this topic can be found in literature. Because of its
importance these works appear in many different research fields.
We quote~\cite{rossi,creutz,sexton} as examples of papers belonging to the
lattice field theory literature. However, similar investigations
could be find for instance in~\cite{yoshida} in a different context.
An useful review for the reader may be~\cite{symplectic} where specific
features of the problem
are discussed including~$(i)$~non separable
Hamiltonians~$(ii)$~situations in which the computation
of the derivatives of $S(q)$ is not expensive.

We want to review shortly the techniques needed to build
efficiently higher order
approximation schemes.
Let us consider a separable Hamiltonian $H = T(p) + V(q)$ with $N$
degrees of freedom.
The symplectic 2-form
$\omega^2 = dp\wedge dq$ and its exterior powers $(\omega^2)^k$
with $k=1\cdots N$ define the
Poincar\`e invariant integrals which give invariant quantities
when integrated;
Liouville theorem is obtained with $k=N$. We want to define a numerical
map of initial conditions
$(q_0, p_0)$ such that $\omega^2$ is exactly preserved and the flow of $H$
is approximated in some
sense\footnote{
We remark that canonical integrators are intrinsically more stable than
the non canonical ones, but
they do not solve completely the problem of numerical stability.
One would be tempted to apply KAM theory to conclude that a conserved
new hamiltonian exists in the
numerical evolution of the integrator. However, KAM theory does not
apply here; as we shall see
we are dealing with a Hamiltonian with discontinuous changes in time.
Indeed, in $N\ge 2$ degrees
of freedom a large numerical diffusion may be observed
(Arnold's diffusion) even with canonical
integrators. One usually expects this effect to be small, but in
general there is no
hope to catch the long time dynamics of $H$.}.

Let $(q(\epsilon), p(\epsilon))$ the exact evolution of $(q_0, p_0)$
under $H$.
A symplectic integrator of order $n$ is defined to be a canonical
transformation $\Phi$
\begin{equation}
(\tilde{q}_0, \tilde{p}_0)\stackrel{\Phi}{\to} (q, p, \epsilon)
\end{equation}
such that
\begin{equation}
\label{symp:def}
(q(\epsilon), p(\epsilon))\stackrel{\Phi^{-1}}{\to}
(\tilde{q}_0, \tilde{p}_0) = (q_0, p_0)
+ O(\epsilon^n).
\end{equation}
Finding symplectic integrators is easy because of the following
observation. Consider a canonical
transformation $(q,p)\to(\tilde{q}_0, \tilde{p}_0)$ such that
$(\tilde{q}_0, \tilde{p}_0)\to(q_0, p_0)$ as $\epsilon\to 0$ and
that the new Hamiltonian is
\begin{equation}
\label{zeroH}
H(\tilde{q}_0, \tilde{p}_0) = \sum_{k=n}^\infty h_k(\tilde{q}_0,
\tilde{p}_0) \epsilon^k
\end{equation}
then the Hamiltonian evolution of $(\tilde{q}_0, \tilde{p}_0)\to(q_0, p_0)$
is zero up to order
$O(\epsilon^{n+1})$ and Eq.~(\ref{symp:def}) holds.
Consider now the chain of canonical transformations
\begin{equation}
(q_l, p_l)\stackrel{K_l}{\to} (q_{l-1}, p_{l-1}) \stackrel{K_{l-1}}{\to}
\cdots
\stackrel{K_1}{\to}  (q_0,p_0),
\end{equation}
where the generating functions are
\begin{equation}
\label{K}
K_i(q_{i-1}, p_i, \epsilon) = -q_{i-1} p_i - \left[
a_i T(p_i) + b_i V(q_{i-1})
\right].
\end{equation}
The explicit transformations are
\begin{eqnarray}
q_i &=& -\partial_{p_i} K_i = q_{i-1} + \epsilon a_i
\partial_{p_i} T(p_i), \nonumber \\
p_{i-1} &=& -\partial_{q_{i-1}} K_i = p_i + \epsilon b_i
\partial_{p_{i-1}} V(q_{i-1}), \\
H_{i-1}(q_{i-1}, p_{i-1}) &=& H_i + \partial_\epsilon K_i .\nonumber
\end{eqnarray}
Imposing the validity of Eq.~(\ref{zeroH}) we obtain the desired symplectic
integrators. We remark that the number of steps $l$ grows faster
than the order $n$ like in standard
Runge-Kutta schemes and finding higher order schemes by brute force
is not possible.
For the actual construction of symplectic integrators a different
approach, namely the operatorial
one, is more practical.
Introducing the notation
\begin{equation}
z = (q, p) \qquad D_G F = \{ F, G \} = \sum_i\left(
\frac{\partial F}{\partial q_i}\frac{\partial G}{\partial p_i}-
\frac{\partial F}{\partial p_i}\frac{\partial G}{\partial q_i}
\right),
\end{equation}
the Hamilton equations are
\begin{equation}
\dot{z} = D_H\ z\quad \Rightarrow\quad z(\epsilon) =
\Phi_{\rm exact}(\epsilon)\ z(0),
\end{equation}
where
\begin{equation}
\Phi_{\rm exact}(\epsilon) = \exp[\epsilon(A+B)]\qquad
A = D_{T(p)} \qquad B = D_{V(q)}
\end{equation}
and we have expressed the Hamiltonian evolution in terms of the
exponential map of the sum of two
differential (non commutative) operators.
The explicit expression of $\Phi_{\rm exact}$ is not known in general.
Again, an integrator of order $n$ is a canonical approximation
$\Phi_n(\epsilon)$ of $\Phi_{\rm exact}$
exact to order $O(\epsilon^n)$.
Recalling Eq.~(\ref{K}) we choose $\Phi$ in the factorized form
\begin{equation}
\label{Tdef}
\Phi(\epsilon) = \prod_{i=1}^{k} e^{\epsilon c_i A}\ e^{\epsilon d_i B}.
\end{equation}
The associated map $z(0)\to z(\epsilon)$ corresponds as before to the
chain of
infinitesimal canonical transformations
\begin{eqnarray}
q_i &=& q_{i-1} + \epsilon c_i \partial_{p_{i-1}}T(p_{i-1}), \\
p_i &=& p_{i-1} - \epsilon d_i \partial_{q_{i-1}}V(q_{i-1}).
\end{eqnarray}
The unknown constants $\{c_i\}$ and $\{d_i\}$ may of course be
determined by brute force.
However, a key observation which simplifies
computation is that for any two operators $A$ and $B$ we have
\begin{equation}
\exp \epsilon A\ \exp \epsilon B\ \exp \epsilon A =
\exp\left\{\epsilon(2A+B)+\frac{\epsilon^3}{6}([B,[B,A]]-[A,[A,B]])+
O(\epsilon^5)\right\}.
\end{equation}
If $\Phi$ satisfies the time reversibility equation
\begin{equation}
\label{timerev}
\Phi(-\epsilon) = \Phi^{-1}(\epsilon),
\end{equation}
then $\Phi$ does not have even terms
\begin{equation}
\Phi(\epsilon) = \exp\left\{
\sum_{k=1}^\infty \epsilon^k X_k
\right\} \ \ \mbox{with}\ \ X_2 = X_4 = \cdots = 0 .
\end{equation}
Using this property we can proof that given a symplectic integrator of
order $2n$ satisfying
Eq.~(\ref{timerev}), the integrator
\begin{equation}
\label{creutz}
\Phi_{2n+2}(\epsilon) = \Phi_{2n}(z_1\epsilon) \Phi_{2n}(z_0\epsilon)
\Phi_{2n}(z_1\epsilon)
\end{equation}
is symplectic, satisfies Eq.~(\ref{timerev}) and is of order $2(n+1)$
if and only if
\begin{equation}
z_0 = -\frac{2^{1/(2n+1)}}{2-2^{1/(2n+1)}}\qquad z_1 =
\frac{1}{2-2^{1/(2n+1)}}.
\end{equation}
Since $\Phi_2$ exists and is obtained using
\begin{equation}
\label{simplest}
k=2\qquad c_1=c_2=\frac{1}{2}\qquad d_1 = 1
\end{equation}
in Eq.~(\ref{Tdef}),  we can show inductively that a
reversible symplectic integrator of arbitrarily high
order $2n$ exists which involves $3^{n-1}$ $\Phi_2$ factors,
namely $3^{n-1}+1$ elementary steps.

We remark that when high order integrators are needed,
more direct approaches can save time. As an example, in~\cite{yoshida}
we find the proposal
\begin{equation}
\Phi^{(m)} = \Phi_2(\epsilon w_m)\cdots \Phi_2(\epsilon w_0)\cdots
\Phi_2(\epsilon w_m).
\end{equation}
By a straightforward computation one finds that $m=3$ and $m=7$ are
enough for the 6th and 8th
order integrator. The constants $\{w_i\}$ are given as numerical solutions of
a set of algebraic equations. The reduced number of steps required
is $8$ and $16$ in this
case, whereas the scheme of Eq.~(\ref{creutz}) needed $10$ and $28$
steps respectively.
\section{P\lowercase{roof of} E\lowercase{q.~(\ref{db1})}}
\label{app1}
We write
\begin{equation}
W(q^{'},p^{'}) P(q^{'},p^{'}\to q^{''},p^{''}) = X_1 + X_2 + X_3
\end{equation}
with
\begin{eqnarray}
X_1 &=& W(q^{'},p^{'}) \int  \Pi(p^{'}\to p_1) K(q^{'},p_1\to q^{''},p_2)
\Pi(p_2\to p^{''})\ dp_1\ dp_2 ,\\
&& \nonumber \\
X_2 &=& W(q^{'},p^{'}) \delta(q^{'}-q^{''}) \Pi^{(2)}(p^{'}\to-p^{''}), \\
&& \nonumber \\
X_3 &=& -W(q^{'},p^{'})\int \Pi(p^{'}\to p_1) K(q^{'},p_1\to q_1,p_2)
\times \label{X3} \\
&& \nonumber \\
&& \times \Pi(-p_1\to p^{''})\delta(q^{'}-q^{''})\ dp_1\ dp_2\ dq_1 ,
\nonumber
\end{eqnarray}
where
\begin{equation}
\Pi^{(2)}(p_1\to p_2) = \int \Pi(p_1 \to p) \Pi(p\to p_2)\ dp .
\end{equation}
Rewriting $X_1$ as
\begin{equation}
X_1 = W(q^{''},p^{''}) \int  \Pi(p^{''}\to p_2) K(q^{''},-p_2\to q^{'},-p_1)
\Pi(p_1\to p^{'})\ dp_1\ dp_2
\end{equation}
it is clear that
\begin{equation}
X_1(q^{'},p^{'}\to q^{''},p^{''}) = X_1(q^{''},-p^{''}\to q^{'},-p^{'}).
\end{equation}
The same propery holds for $X_2$ since $\Pi^{(2)}$ satisfies Detailed Balance
like $\Pi$ does. Finally
\begin{eqnarray}
\lefteqn{X_3(q^{'},p^{'}\to q^{''},p^{''}) =} && \nonumber\\
&&
-\int dq_1\ dp_1\ dq_2\ dp_2\ \delta(q_2-q^{'})\delta(q_2-q^{''})
\Pi(p_1\to p^{'}) \Pi(p_1\to -p^{''}) \times \\
&&
\times K(q_1,-p_2\to q_2,-p_1)
W(q_1,p_2) \nonumber
\end{eqnarray}
from which it is evident that
\begin{equation}
X_3(q^{'},p^{'}\to q^{''},p^{''}) = X_3(q^{''},-p^{''}\to q^{'},-p^{'}).
\end{equation}
\section{P\lowercase{roof of} E\lowercase{q.~(\ref{db2})}}
\label{app2}
We now proof Eq.~(\ref{db2}). We write as before
\begin{equation}
W(q^{'},p^{'},y^{'}) P(q^{'},p^{'},y^{'}\to q^{''},p^{''},y^{''}) =
Y_1 + Y_2 + Y_3
\end{equation}
then
\begin{eqnarray}
Y_1 &=& W(q^{'},p^{'},y^{'})
\int \Pi(y^{'}\to y_1) T_{rot}(y_1,p^{'}\to y_2 ,p_1)
K(q^{'},p_1\to q^{''},p_2)  \times \nonumber\\
&&
\times T_{rot}(y_2,p_2\to y_3,p^{''})\Pi(y_3\to y^{''})\
dp_1\ dp_2\ dy_1\ dy_2\ dy_3
\end{eqnarray}
and we see that
\begin{equation}
Y_1(q^{'},p^{'},y^{'}\to q^{''},p^{''},y^{''}) = Y_1(q^{''},-p^{''},y^{''}
\to q^{'},-p^{'},y^{'}).
\end{equation}
Regarding $Y_2$
\begin{eqnarray}
Y_2 &=& W(q^{'},p^{'},y^{'}) \delta(q^{'}-q^{''})
\int \Pi(y^{'}\to y_1) T_{rot}(y_1,p^{'}\to y_2 ,p_1)
T(q^{'},p_1\to q_1,p_2)  \times \nonumber \\
&&
\times T_{rot}(y_2,-p_1\to y_3,p^{''})\Pi(y_3\to y^{''})\
dq_1\ dp_1\ dp_2\ dy_1\ dy_2\ dy_3
\end{eqnarray}
but
\begin{eqnarray}
\lefteqn{\int T_{rot}(y_1,p^{'}\to y_2,p_1)
T_{rot}(y_2,-p_1\to y_3,p^{''})\ dy_2\ dp_1 = } && \nonumber\\
&&
= \int \langle y_3,-p^{''} | e^{\epsilon H_{rot}} | y_2,p_1 \rangle
\langle  y_2,p_1 | e^{-\epsilon H_{rot}} | y_1,p^{'} \rangle \
dy_2\ dp_1 = \\
&&
=\langle y_3,-p^{''} | y_1,p^{'} \rangle = \delta(y_1-y_3)
\delta(p^{'}+p^{''})\nonumber
\end{eqnarray}
whence
\begin{equation}
Y_2 = W(q^{'},p^{'},y^{'})\delta(q^{'}-q^{''})
\delta(p^{'}+p^{''})\Pi^{(2)}(y^{'}-y^{''})
\end{equation}
and again
\begin{equation}
Y_2(q^{'},p^{'},y^{'}\to q^{''},p^{''},y^{''}) =
Y_2(q^{''},-p^{''},y^{''}\to q^{'},-p^{'},y^{'}).
\end{equation}
Finally,
\begin{eqnarray}
Y_3 &=& W(q^{'},p^{'},y^{'})\delta(q^{'}-q^{''})
\int \Pi(y^{'}\to y_1) T_{rot}(y_1,p^{'}\to y_2 ,p_1)
K(q^{'},p_1\to q_1,p_2)  \times \nonumber \\
&&
\times T_{rot}(y_2,-p_1\to y_3,p^{''})\Pi(y_3\to y^{''})
\ dq_1\ dp_1\ dp_2\ dy_1\ dy_2\ dy_3 = \\
&&
=\delta(q^{'}-q^{''})
\int \Pi(y_1\to y^{'}) \Pi(y_3\to y^{''}) T_{rot}(y_1,p^{'}\to y_2 ,p_1)
T_{rot}(y_2,-p_1\to y_3,p^{''})\times \nonumber \\
&&
K(q_1,-p_2\to q^{'},-p_1)  W(q^{'},p_1,y_2)
\ dq_1\ dp_1\ dp_2\ dy_1\ dy_2\ dy_3. \nonumber
\end{eqnarray}
Since in the motion
determined by $H_{rot}$ we can view $y$ as a coordinate it follows that
\begin{equation}
T_{rot}(y^{'},p^{'}\to y^{''},p^{''}) =
T_{rot}(y^{''},-p^{''}\to y^{'},-p^{'})
\end{equation}
We can verify that changing $p^{'}$ and $p^{''}$ into $-p^{''}$, $-p^{'}$
and exchanging the names of $y_1$ and $y_3$ the expression for $Y_3$
does not change.

N. Madras and A.D. Sokal,
Jour. of Stat. Phys. {\bf 50}, 109 (1988).
\bibitem{kramers}
H. A. Kramers, Physica {\bf 7}, 284 (1940)
\bibitem{risken}
H. Risken, {\sl The Fokker-Planck Equation}, Eds. Springer-Verlag, (1984).
\bibitem{drummond}
I.T. Drummond, S. Duane and R. R. Horgan,
Nucl. Phys. B{\bf 220}, 119 (1983).
\bibitem{chin}
S. A. Chin,
Nucl. Phys. (Proc. Suppl.) B{\bf 9}, 498 (1989).
\bibitem{rossky}
P. J. Rossky, J. D. Doll and H. L. Friedman,
J. Chem. Phys. {\bf 69}, 4628 (1978).
\bibitem{horowitz1}
A. M. Horowitz,
Phys. Lett. {\bf 156}B, 89 (1985).

A. M. Horowitz,
Nucl. Phys. B{\bf 280}, 510 (1987).
\bibitem{horowitz2}
A. M. Horowitz,
Phys. Lett. {\bf 268}B, 247 (1991).
\bibitem{menotti}
P. Menotti and E. Onofri,
Nucl. Phys. B{\bf 190}, 288 (1981).
\bibitem{hmc1}
S. Duane, A. D. Kennedy, B.J. Pendleton and D. Rowan,
Phys. Lett. {\bf 195} B, 216 (1987).
\bibitem{hmc2}
M. Creutz,
Phys. Rev. D{\bf 38}, 1228 (1988).

R. Gupta, G. W. Kilcup and S. R. Sharpe,
Phys. Rev. D{\bf 38}, 1278 (1988).

S. Gupta, A. Irback, F. Karsch and B. Petersson,
Phys. Lett. {\bf 242}B, 437 (1990).
\bibitem{rossi}
P. Marenzoni, L. Pugnetti and P. Rossi,
Parma University preprint, June 1993.
\bibitem{rossi}
M. Campostrini and P. Rossi,
Nucl. Phys. B{\bf 329}, 753 (1990).
\bibitem{creutz}
M. Creutz and A. Gocksch,
Phys. Rev. Lett. {\bf 63}, 9 (1989).
\bibitem{sexton}
J. C. Sexton, D. H. Weingarten,
Nucl. Phys. (Proc. Suppl.) B{\bf 26}, 613 (1992).
\bibitem{yoshida}
H. Yoshida,
Phys. Lett.A{\bf 150}, 262 (1990).
\bibitem{symplectic}
R. I. McLachlan and P. Atela,
Nonlinearity {\bf 5}, 541 (1992).
\begin{table}
\caption{Some admissible $\gamma$, $\theta_i$ for the $(q,p,y)$ scheme}
\vskip 0.5truecm
\begin{tabular}{cccccccc}
\label{solutions}
N & $\lambda$ & $\gamma$ & $\theta_1$ & $\theta_2$ &
 $\theta_3$ & $\theta_4$ & $\theta_5$ \\
\tableline
1 & $-\sqrt{3}$ & $3\sqrt{3}$ & $2\sqrt{2}$          & $-$ & $-$
        & $-$ & $-$\\
2 & $-\sqrt{5}$ & $4\sqrt{5}$ & $2$                  & $5$ & $-$
        & $-$ & $-$\\
3 & $-\sqrt{7}$ & $5\sqrt{7}$ & $4/\sqrt{5}$         & $7/\sqrt{5}$
       & $\sqrt{56}$ & $-$
& $-$  \\
4 & $-3$        & 18          & $2\sqrt\frac{5}{7}$  & $\sqrt\frac{243}{35}$
 &
 $\sqrt\frac{96}{5}$ & $\sqrt{105}$ & $-$  \\
5 & $-\sqrt{11}$        & $7\sqrt{11}$          & $\sqrt\frac{8}{3}$
  & $\frac{11}{\sqrt{21}}$ &
 $\sqrt\frac{88}{7}$ & $\sqrt{33}$ & $4\sqrt{11}$ \\
\end{tabular}
\end{table}

\references
%
\bibitem[*]{e1} beccaria@hpth4.difi.unipi.it, beccaria@vaxsns.sns.it
\bibitem[*]{e2} curci@mvxpi1.difi.unipi.it
%
\bibitem{beccaria}
M. Beccaria, G. Curci and A. Vicer\'e,
Phys. Rev. E (accepted for publication).
%
\bibitem{sokal}
A. Sokal,
Monte Carlo Methods in Statistical mechanics:
Foundations and New Algorithms, in
{\sl Cours de Troisi\`eme Cycle de la Physique en Suisse Romande}.
\end{document}